\documentclass[twoside]{dis09}
\usepackage[latin1]{inputenc}
\usepackage[dvips]{graphicx,epsfig,color}
\usepackage{wrapfig,rotating}
\usepackage{amssymb,amsmath,array}
\usepackage{epsfig,dcolumn}
\usepackage{graphicx}
\usepackage{bm}
\usepackage{amsfonts}
\usepackage{amsmath}
\usepackage{amssymb}
\usepackage{graphicx}

\def\lsim{\mathrel{\rlap{\lower4pt\hbox{\hskip1pt$\sim$}}
    \raise1pt\hbox{$<$}}}
\def\gsim{\mathrel{\rlap{\lower4pt\hbox{\hskip1pt$\sim$}}
    \raise1pt\hbox{$>$}}}

\newcommand{\be}{\begin{equation}}
\newcommand{\ee}{\end{equation}}
\newcommand{\ba}{\begin{eqnarray}}
\newcommand{\ea}{\end{eqnarray}}

\def\wtilp{{{\widetilde p}}}

\def\qs{\displaystyle{\not} q }
\def\ks{\displaystyle{\not} k }

\begin{document}
\title{Reggeon Non--Factorizability \\ and the $J=0$ Fixed Pole in DVCS}

\author{Stanley J. Brodsky$^1$, Felipe J. Llanes-Estrada$^2$,
J. Timothy Londergan and Adam P. Szczepaniak$^3$
%
\vspace{.3cm}\\
%
1- SLAC National Laboratory, Stanford University\\
2575 Sand Hill Road,  94025 Menlo Park, CA, USA
%
\vspace{.1cm}\\
2- Universidad Complutense de Madrid, Departamento de 
F\'{\i}sica Te\'orica I
\\ Avda. Complutense s/n, 28040 Madrid, Spain
\vspace{.1cm}\\
3- Indiana University, Nuclear Theory Center and Physics Department\\
Bloomington, IN 47405, USA.
}

\maketitle

\begin{abstract}
We argue that deeply virtual Compton scattering will display Regge  
behavior $\nu^\alpha_R(t)$ at high energy at fixed-$t$,  even at high photon virtuality,
not necessarily conventional scaling. A way to see this is to track the Reggeon contributions to   
quark-nucleon scattering and notice that the resulting Generalized Parton 
Distributions would have divergent behavior at the break--points. 
In addition,  we show that the direct two-photon to quark coupling will be accessible at large $t$ where it dominates the DVCS amplitude for large energies. 
This contribution, the $J=0$ fixed--pole, should be part of the future DVCS experimental 
programs at Jlab or LHeC.
\end{abstract}

It is commonly believed that  the DVCS (Deeply Virtual Compton Scattering) amplitude scales;  i.e., at high energy, its  energy $\nu=(s-u)/4$. and photon virtuality $Q^2$ dependence enters only through the scaling variable $\xi=Q^2/(2\nu)$.
However, we have argued\footnote{Talk presented by FJLE at DIS09, Madrid, April 
$29^{\rm th}$ 2009, see slides \cite{urltalk}. Manuscript number 
SLAC-PUB-13694; arXiv:0906.5515 .}
that instead, DVCS and  possibly other hard exclusive processes will have the characteristic Regge behavior $T\propto \nu^{\alpha_R(t)}$ at large $s$ at fixed $t$ and photon virtuality $Q^2$.
This distinction has been noted in the past (see, for example, the work  of Bjorken and Kogut~\cite{Bjorken:1973gc} and the 
 $t$--channel analysis in ref.~\cite{Szczepaniak:2006is}.   Here we  address the consequences of Regge behavior for  Generalized Parton Distributions (GPD's).
\footnote{The origin of Regge behavior in forward Compton scattering and deep inelastic scattering based on  the handbag diagram dates back to the covariant parton model of Landshoff, Polkinghorne and Short~\cite{Landshoff:1970ff}. The analysis was extended to virtual Compton scattering in ref. ~\cite{Brodsky:1973hm}.}
 
\begin{figure}
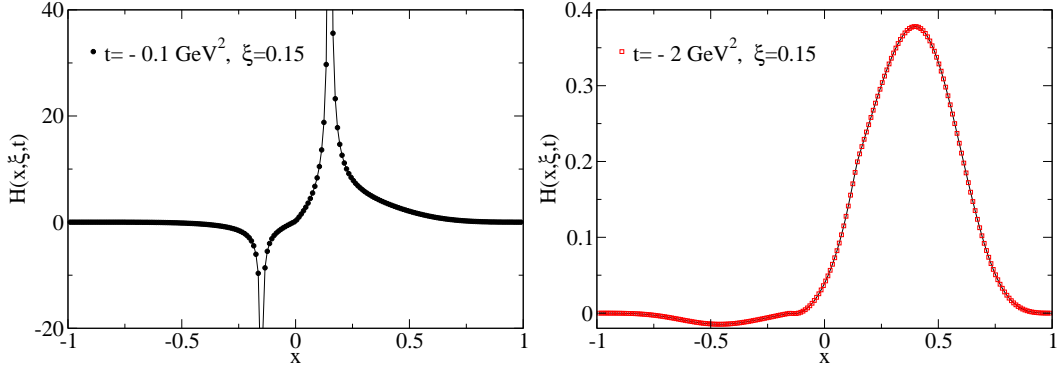

\includegraphics[height=1.9in]{Llanes_Estrada_Felipe.fig1.eps}
\includegraphics[height=1.9in]{Llanes_Estrada_Felipe.fig2.eps}
\caption{\label{fig:gpdplot}
Left panel: low--t Generalized Parton Distributions may present Regge 
divergences at the break--points $x=\pm\xi$, $H(x,\xi,t)\propto 
(x-\xi)^{-\alpha(t)}$. 
Right panel: high--t GPD's are the sum of an analytic part plus a 
non--analytic part that vanishes at the break--points. Collinear 
factorization then holds, in a regime where $Q^2>>-t>M_N^2$.
}
\end{figure}

Conventional analyses of DVCS are based on the collinear factorization theorem~\cite{Collins:1996fb} which expresses the Compton amplitude  in terms of the convolution of hard--scattering kernel and a soft hadronic amplitude,  which for the helicity conserving case,  is the $H$ GPD~\cite{Ji:1998pc},
\begin{equation}
\label{factorizedDVCS}
T_+(\xi,t)= - \int_{-1}^1 dx H(x,\xi,t) \left[ \frac{1}{x+\xi-i\epsilon} +\frac{1}{x-\xi+i\epsilon}
\right]\
\end{equation}
This expression is valid whenever the  GPD $H$ is continuous at the ''break--points''  $x=\pm \xi$. This continuity is an assumption in the derivation of the factorization theorem~\cite{Collins:1996fb}.
In the handbag diagram, the longitudinal momentum fraction of the extracted and returning quark are respectively $k^+/P^+ = x-\xi$ and $x+\xi$.  Thus the break--points correspond to either quark having zero momentum fraction and infinite light cone energy. Thus the parton-nucleon amplitude  become singular at the break-points if it has energy  dependence  suggested  by Regge scattering. 
Indeed in  the case of DIS, it is well--known that structure functions have a strong Regge divergence $F_2(x)\propto x^{-\alpha(0)}$ when the probed quark has zero 
 momentum fraction at $x=0$. It is thus not unnatural to expect the same for $H$ in exclusive processes when $x\to \pm \xi$.  DVCS additionally depends on the squared momentum transfered to the proton,
 $-t$.  Reggeons are known to have $\alpha_R(t)>0$ for small $-t$ and $\alpha_R(t)<0$ for sizeable $-t>M_N^2$~\cite{Coon:1974wh}. Therefore, Eq.(\ref{factorizedDVCS}) is expected to hold  for some large $-t$ where the GPD becomes finite, but may not be  applicable to small $-t$. The situation is depicted in figure \ref{fig:gpdplot}.

In addition to intuitive arguments, we have formally derived the break--point divergent behavior $H(x,\xi,t)\propto (x\pm \xi)^{-\alpha}$ in~\cite{Szczepaniak:2007af}.  We now briefly recall the demonstration.
The hadron part of the Compton amplitude is a spin--tensor, the matrix element of a product of currents
\begin{equation}
T^{\mu\nu}  = i  \int d^4z e^{i\frac{q'+q}{2} z} \langle p'\lambda' | 
  T J^\mu(z/2) J^\nu(-z/2) | p\lambda\rangle.
\label{Wdef}
\end{equation} 
For large $Q^2$ the $z$-integral peaks at $z^2 \sim~1/Q^2$ and using the 
leading order operator product expansion of QCD we replace the product of the 
two currents by a product of two quark field operators and a free propagator 
between the photon interaction points $(z/2,-z/2)$  
 \begin{eqnarray}
 T^{\mu\nu}  =-i e_q^2\int \frac{d^4k}{(2\pi)^4} A_{\beta\alpha}(k,\Delta,p,\lambda,\lambda'). \left\{ 
   \frac{ \left[\gamma^\mu \left(\ks + \frac{\qs+\qs'}{2} \right)
  \gamma^\nu\right]_{\alpha\beta}}{\left( \frac{q+q'}{2} + 
  k \right)^2 + i\epsilon} - 
    \frac{ \left[\gamma^\nu \left(-\ks + \frac{\qs+\qs'}{2} \right)
  \gamma^\mu\right]_{\alpha\beta}}{\left( \frac{q+q'}{2} - 
 k \right)^2+i\epsilon} \right\} \label{bigT}   
\end{eqnarray} 
Here $A$ is the parton-nucleon scattering amplitude~\cite{Brodsky:2005vt} with quark propagators included  
\begin{eqnarray} 
   A_{\beta\alpha} = 
   -i \int d^4z e^{-ikz} \langle p'\lambda'|T \bar\psi_\alpha(z/2) 
  \psi_\beta(-z/2) | p\lambda\rangle. 
\label{Adef} 
\end{eqnarray}
This is a hadron-hadron amplitude, and as is the case of  $NN$, $N\pi$, $\pi\pi$ scattering. it is expected to have Regge behavior~\cite{Brodsky:1973hm}. For illustration purposes, from the several spin Dirac matrices contributing, it is sufficient to  pick up one, {\it i.e.} corresponding to $t$-channel  scalar exchange, 
 $A \propto   \frac{\ks_{\beta\alpha}}{4}  \delta_{\lambda'\lambda}$.  As shown in~\cite{Szczepaniak:2007af}, the forward Compton amplitude $\gamma^* p\to \gamma^* p$ in Eq.~(\ref{bigT}) reproduces the known DIS properties.
However, when applied to DVCS $\gamma^* p \to \gamma p$,  it yields a Compton amplitude of the form
\begin{equation} 
 T^{\mu\nu}_+ = -\delta_{\lambda'\lambda} e_q^2  
\left[ n^\mu \wtilp^\nu  + n^\nu \wtilp^\mu - g^{\mu\nu} (n \cdot \wtilp) \right] 
 \left[ 
  \left(  \frac{Q^2} {\xi \mu^2 } \right)^{\alpha^+_s} F^+_s(\xi) + 
 \left( \frac{Q^2} {\xi \mu^2}\right)^{\alpha^+_u} F^+_u(\xi) \right] \ .
\label{notscaling}
\end{equation}
From the GPD point of view, this arises because $H$ is
a projection of the parton-nucleon amplitude 
$H(x,\xi,t) \propto p^+ \int \frac{d^4k}{(2\pi)^4}
\delta\left(x p^+ -k^+\right) A$, so the Regge behavior of $A$ (natural for a hadron-hadron scattering amplitude) is inherited by the GPD in 
Eq.~(\ref{factorizedDVCS}) \footnote{A different issue is the inadequacy of the handbag diagram to represent leading-twist diffractive DIS, even in light--cone gauge\cite{Brodsky:2002ue}.}.

\begin{figure}
\includegraphics[height=2.5in]{Llanes_Estrada_Felipe.fig3.eps}
\includegraphics[height=1.2in]{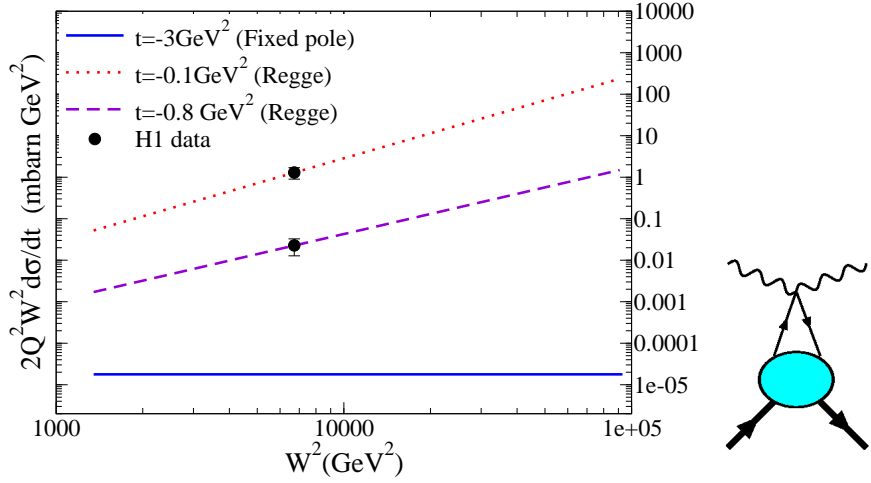}
\caption{\label{fig:compton}
DVCS differential cross section as a function of the energy for constant scaling variable $\xi$ fixed by the H1 kinematics~\cite{Aktas:2005ty}, 
$Q^2=8\ GeV^2$, $W=82\ GeV$.
We argue that the DVCS amplitude at large $s$, $Q^2$  is
Regge--behaved. This implies scaling violations which become less 
prominent for larger $t$. (The quantity plotted, $2Q^2W^2d\sigma/dt$ should scale at LO.)
When $t$ is large enough such that $\alpha(t)<0$,
all Reggeons have receded. The resulting amplitude at large $s$ is then a real, 
$Q^2$--independent constant, the $J=0$ fixed pole, which reveals the 
two--photon coupling of quarks (right).
}
\end{figure}

However, Eq.~(\ref{factorizedDVCS}) and (\ref{notscaling}) come together at large $-t$. In that case $\alpha_R(-t)<0$ for conventional Reggeons and their contribution to the amplitude decreases with $s$. But a contribution remains with $\alpha_R=0$, independent of $t$ and photon virtuality; it dominates the amplitude at sizeable $-t$, turning it to a real constant:  the $J=\alpha=0$ fixed pole (see fig. \ref{fig:compton}). 
This contribution is familiar in atomic physics, corresponding to high energy Compton scattering on the atomic electrons.
This constant acts as a subtraction in the dispersion relation for the Compton amplitude\cite{Brodsky:2008qu}
\be
\label{dispDVCS} 
T_+(\xi,t)=C_\infty(t)+\frac{\xi^2}{\pi}
\int_0^1\frac{2dx}{x}\frac{{\rm Im}\ T_+(x,t)}{\xi^2-x^2-i\epsilon}\ .
\ee
The value of $C_\infty$ is not fixed by the dispersive integral, and its necessity is revealed by the dynamical insight that the target nucleon's components which carry charge are elementary, so that a seagull-like $\gamma\gamma qq$ coupling is active\footnote{A subtraction term is needed for Compton scattering since the low energy amplitude is fixed by the negative--signed Thomson term; the unsubtracted dispersion relation has the wrong sign. $C_{\infty}$ receives contributions from the Thomson term and the constant contribution from the dispersive integral, which curiously enough, Damashek and Gilman\cite{Gilman} found to be small.}. 

The $J=0$ contribution arises from a term in the numerator of the Feynman propagator in the case of spin-1/2 quarks which cancels the energy denominator; it thus has no imaginary part and no $s$ dependence (hence its real, constant nature). The resulting seagull--like interaction extends over $t-z$, conjugate to $k^+$ and $k^{'+}$. This gives it a characteristic $\gamma^+/x$ dependence:
\begin{eqnarray} 
\frac{\not k + \not q + m}{(k+q)^2-m^2+i\epsilon} \to
\frac{\gamma^+}{2p^+}\left( \mathbf{\frac{1}{x}} +
\frac{\xi}{x}\frac{1}{x-\xi+i\epsilon} \right)  
= \frac{\gamma^+}{2p^+} \frac{1}{x-\xi+i\epsilon}
 \nonumber \\
- \frac{  \not k - \not q' + m}{(k-q')^2-m^2+i\epsilon} \to
\frac{\gamma^+}{2p^+}\left( \mathbf{\frac{1}{x}} -
\frac{\xi}{x}\frac{1}{x+\xi-i\epsilon} \right) 
= \frac{\gamma^+}{2p^+} \frac{1}{x+\xi-i\epsilon} \ . \label{fixedpoleinprops}
\end{eqnarray}
Comparing Eq.(\ref{factorizedDVCS}), ~(\ref{dispDVCS}) we find at high enough energy
\be \label{subtconstantDVCS}
C_\infty(t)=\lim_{\xi\to 0} T_+(\xi,t)= -2\int_{-1}^1 dx \frac{H(x,0,t)}{x}
=-2 F_{1/x}(t)
 \ .
\ee
The real part of the DVCS amplitude can be  identified  by interference with the Bethe-Heitler amplitude, thus  accessing the $1/x$ form factor of the nucleon~\cite{Brodsky:2007fr}.
In our recent work~\cite{Brodsky:2008qu}, we have been able to establish how the same fixed pole amplitude also appears in real\footnote{Hints of the fixed pole behavior can be seen in the Cornell and Jefferson lab data~\cite{Shupe:1979vg,Danagoulian:2007gs}.} as well as doubly virtual Compton scattering, showing its universal character. This extension of the $1/x$ form factor to small $-t$  is achieved by analytic continuation in $t$, defining a valence part that is free of Regge behavior $H_v(x,0,t) \equiv  H(x,0,t) - H_R(x,t)$, 
\ba
H_R(x,t) \equiv \theta(x) \sum_{\alpha>0}\frac{\gamma_\alpha(t)}{x^{\alpha(t)}} - \theta(-x)\sum_{\bar \alpha>0}
\frac{\bar\gamma_{\bar \alpha}(t)}{(-x)^{\bar \alpha(t)}}  
\\
F_{1/x}(t) \equiv  \int_{-1}^1 \frac{dx}{x} H_v(x,0,t) -
\sum_{\alpha>  0} \frac{\gamma_\alpha}{\alpha(t)} -
\sum_{\bar \alpha>  0} \frac{ \bar \gamma_{\bar \alpha}}{\bar \alpha(t)}
\ea
In conclusion, we have argued that  DVCS at fixed $-t$ is Regge--behaved and at large $-t$ one can extract from it a $J=0$ fixed pole contribution which is a distinct part of the handbag diagram for spin $1/2$ constituents.
In the case of spin zero quarks, the handbag diagram is not sufficient to obtain the DVCS amplitude;  the seagull is necessary. This gives an energy independent $J=0$  fixed pole contribution to the DVCS amplitude which is independent of the incident or outgoing photon virtualities at fixed $t$, which is not given directly by the GPD--based handbag diagram.



\section*{Acknowledgments}
Presented by FJLE. Financial support from grants
FPA 2008-00592/FPA, FIS2008-01323, PR27/05-13955-BSCH (Spain),
and DE-FG0287ER40365, NSF-PHY0555232 (USA). 


\end{document}